\documentclass[preprint]{elsarticle}
\usepackage{mathrsfs}
\usepackage{amsmath}
\usepackage[usenames, dvipsnames]{color}
\usepackage{array}
\usepackage{multirow}
\usepackage{subfigure}

\usepackage{geometry}                             
\usepackage{graphicx}
\usepackage{amssymb}
\usepackage{epstopdf}
\usepackage{xargs}                 
\usepackage[pdftex,dvipsnames]{xcolor}  
\usepackage[colorinlistoftodos,prependcaption,textsize=tiny]{todonotes}
\usepackage{siunitx}
\usepackage{graphicx}
\usepackage{caption}
\usepackage{verbatim}
\usepackage{verbatimbox}
\usepackage{xcolor, colortbl}
\usepackage[hidelinks]{hyperref}
\usepackage[utf8]{inputenc}
\usepackage{xfrac}

\newcommandx{\unsure}[2][1=]{\todo[linecolor=red,backgroundcolor=red!25,bordercolor=red,#1]{#2}}
\newcommandx{\change}[2][1=]{\todo[linecolor=blue,backgroundcolor=blue!25,bordercolor=blue,#1]{#2}}
\newcommandx{\info}[2][1=]{\todo[linecolor=OliveGreen,backgroundcolor=OliveGreen!25,bordercolor=OliveGreen,#1]{#2}}
\newcommandx{\improvement}[2][1=]{\todo[linecolor=Plum,backgroundcolor=Plum!25,bordercolor=Plum,#1]{#2}}
\newcommandx{\thiswillnotshow}[2][1=]{\todo[disable,#1]{#2}}

\newcommand{\mbtrack}{{\it mbtrack}}
\newcommand{\mbtrackcuda}{{\it mbtrack-cuda}}

\newcolumntype{C}{>{\centering\arraybackslash}m{1.5cm}}
\newcolumntype{X}{>{\centering\arraybackslash}m{2.5cm}}

\begin{document}

\title{Calculation of Longitudinal Collective Instabilities with {\mbtrackcuda}}

\author{Haisheng Xu\footnote{Haisheng.Xu@ihep.ac.cn, the author is presently in the Institute of High Energy Physics, CAS, Beijing, China}, 
Uldis Locans\footnote{Uldis.Locans@gmail.com}, Andreas Adelmann\footnote{Andreas.Adelmann@psi.ch}, Lukas Stingelin\footnote{Lukas.Stingelin@psi.ch}, 
\\Paul Scherrer Institut, CH-5232 Villigen PSI, Switzerland}

\begin{abstract}
Macro-particle tracking is a prominent method to study the collective beam instabilities in accelerators. 
However, the heavy computation load often limits the capability of the tracking codes. One widely used macro-particle tracking code to 
simulate collective instabilities in storage rings is {\mbtrack}. 
The Message Passing Interface (MPI) is already implemented in the original {\mbtrack} to accelerate 
the simulations. However, many CPU threads are requested in {\mbtrack} for the analysis of the coupled-bunch instabilities.
Therefore, computer clusters or desktops with many CPU cores are needed. Since these are not always available, 
we employ as alternative a Graphics Processing Unit (GPU) with CUDA programming interface to run such simulations 
in a stand-alone workstation. All the heavy computations have been moved to the GPU. 
The benchmarks confirm that {\mbtrackcuda} can be used to analyze coupled bunch instabilities up to at least 484 bunches. Compared to {\mbtrack} on an 
8-core CPU, 36-core CPU and a cluster, {\mbtrackcuda} is faster for simulations of up to 3 bunches. For 363 bunches, {\mbtrackcuda} needs about 
six times the execution time of the cluster and twice of the 36-core CPU. The multi-bunch instability analysis shows that the length of the 
ion-cleaning gap has no big influence, at least at filling to \sfrac{3}{4}.
\end{abstract}

\maketitle

\section{Introduction}

Synchrotron light sources have been a powerful tool for condensed matter physics, material science, biology and medicine since about 1968~\cite{Rowe:1973xw}. 
The demand for higher 
brightness from the users of synchrotron light sources has pushed for improved  
performance of the accelerators, for instance, to reduce the emittance of electron beams further. 
In recent years, there have been remarkable improvements on the design of ultra-low emittance storage rings. 
Thanks to the applications of the concepts of Multi-Bend Achromat (MBA) \cite{DieterMBA93}, Longitudinal 
Gradient Bends (LGBs) \cite{STREUN201598}, and Anti-Bends (ABs) \cite{STREUN2014148}, an emittance of 
the order of 100~pm (or even lower) has been achieved in many new designs of storage 
rings and measured at MAX~IV~\cite{Leemann:2017nis} during early commissioning. All the above mentioned concepts are employed in the design of the storage ring 
for the Swiss Light Source Upgrade (SLS-2) \cite{STREUNIPAC2016}.

Since the lattices of the ultra-low emittance rings need strong focusing provided by the strong quadrupole magnets, 
high-field sextupole magnets are needed for the correction of chromatic and geometric aberration. Vacuum 
chambers with small cross sections are therefore considered in the ultra-low emittance storage rings, causing 
high impedance. Furthermore, ultra-low emittance storage rings are more 
sensitive to the impedance induced instabilities. 

The {\mbtrack} code \cite{Skripka2016221} is a multi-bunch macro-particle tracking code which can be used to 
study both single-bunch and coupled-bunch instabilities in electron storage rings. It has been used in different 
synchrotron light sources, such as, SOLEIL \cite{NAGAOKAPAC09}, MAX~IV \cite{Skripka2016221,KLEINIPAC2013}, etc. The Message Passing Interface (MPI) is implemented 
in {\mbtrack} for acceleration of the multi-bunch tracking. For an n-bunch simulation using {\mbtrack}, 
where n is an integer, (n+1) MPI processes are needed. Therefore, a large scale computing cluster is usually preferred 
to perform the multi-bunch simulations in the storage rings of the synchrotron light sources. For instance, 
391 processes are required for the analysis of the nominal operation mode of Swiss Light Source (SLS) \cite{Milas:2010zz,BusseGrawitz:1999jn}, including 
390 bunches in the bunch train.

In order to reduce the demands for large scale clusters for multi-bunch simulations, we developed a GPU version 
of {\mbtrack} (\mbtrackcuda) \cite{LOCANSTHESIS17,LOCANSIPAC17}, in which the computations are offloaded to a Graphics Processing Unit (GPU). 
Taking advantage of a state-of-the-art GPU, which has a massively parallel architecture including thousands of cores, 
one can parallelize the tracking of all the macro-particles in an efficient manner. An NVIDIA graphics card was chosen as the hardware 
to develop the {\mbtrackcuda}. CUDA \cite{CUDAREF}, which is a parallel computing platform and programming model invented by NVIDIA, 
is used for programming. The resulting {\mbtrackcuda} manages to carry out the multi-bunch simulations in a stand-alone 
workstation equipped with an NVIDIA Tesla K40c GPU.

Recently, the advantages of GPU computing attracts interest in the accelerator physics community for the analysis of collective 
instabilities. For example, the GPU implementation for the {\tt elegant} code \cite{GPUELEGANTIPAC2013,GPUELEGANTIPAC2015} is also 
undergoing and the {\it Inovesa} code \cite{InovesaPRAB}, which is a Vlasov-Fokker-Plank solver, also benefits from modern GPU 
computing. The development of {\mbtrackcuda} provides a powerful alternative, especially for the users of {\mbtrack}.

In this paper, we present the development of the {\mbtrackcuda} code and the studies of longitudinal collective 
instabilities for SLS-2 carried out by this code. The rest of this paper is organized as follows. In Section~\ref{sec:CodeInfo}, 
the detailed information of {\mbtrackcuda} code is presented. The benchmark of the code is presented 
in Section~\ref{sec:CodeBenchmark}. The simulation study of the longitudinal coupled-bunch instability for SLS-2 
by {\mbtrackcuda}, are presented in Section~\ref{sec:CodeSimulations}. The conclusions are in Section~\ref{sec:Conclusion}.

\section{GPU acceleration of {\mbtrack}}
\label{sec:CodeInfo}

\subsection{The introduction of {\mbtrackcuda} development}
The {\mbtrackcuda} code \cite{LOCANSTHESIS17} is an expansion of the original version. The main architecture of the code is kept 
identical to the original version. Similar to {\mbtrack}, {\mbtrackcuda} allows the users to choose the effects 
included in the simulations.

\begin{figure}[htb]
\centering
\subfigure[]{\includegraphics[height=45mm,width=0.45\linewidth]{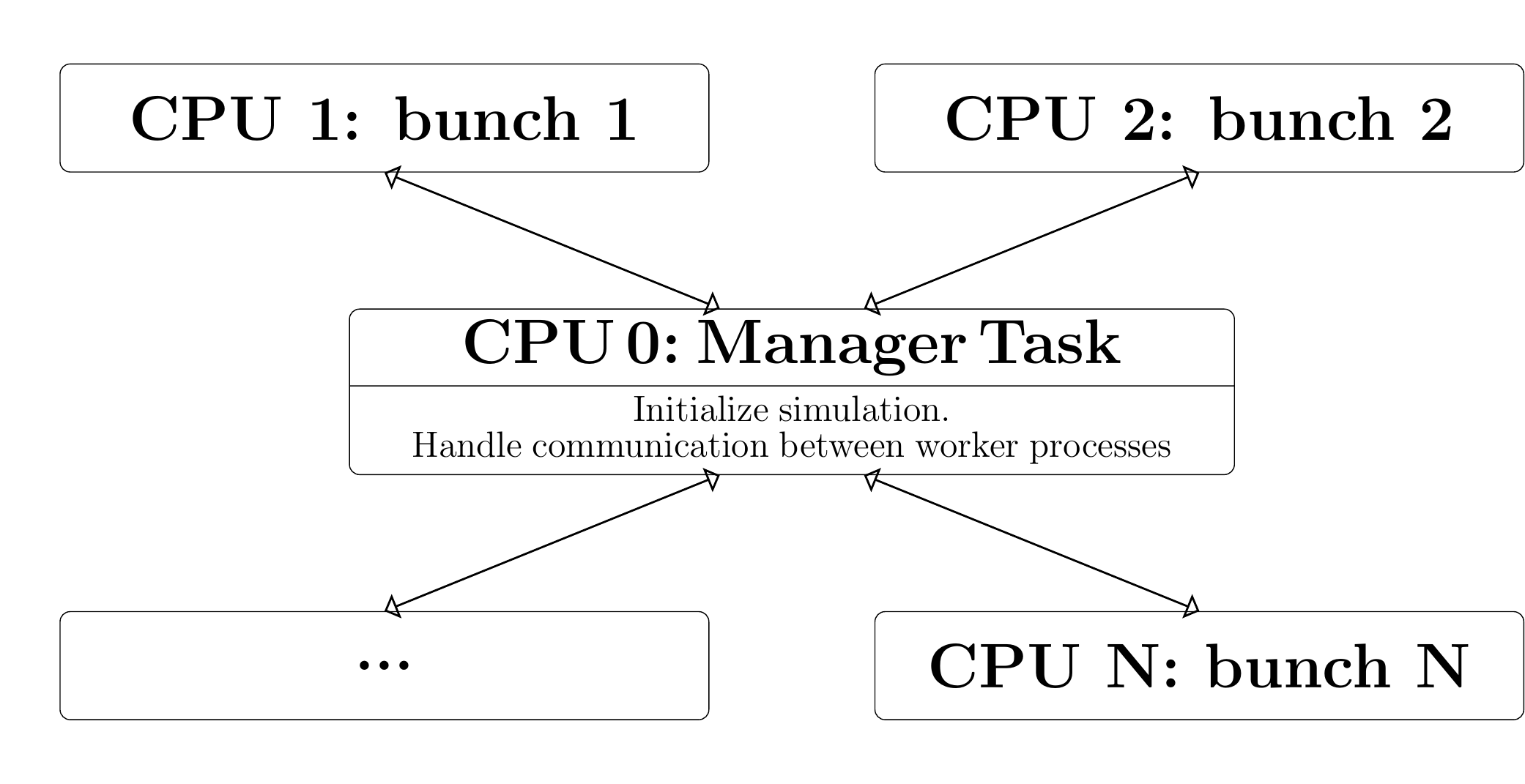}}
\subfigure[]{\includegraphics[height=42mm,width=0.45\linewidth]{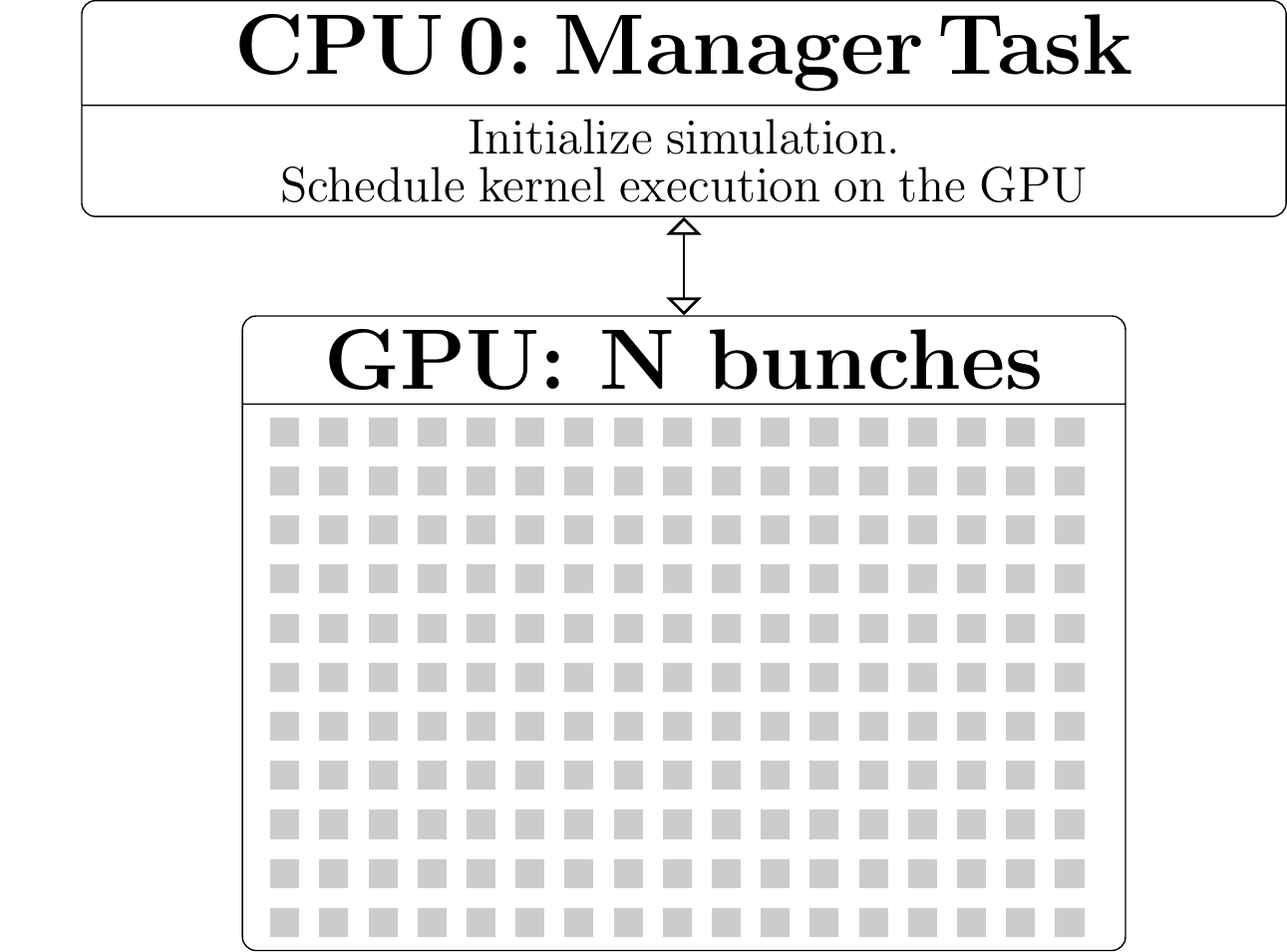}}
\caption{\label{fig:mbtrack-parallelize}Parallelization in the original {\mbtrack} and the {\mbtrackcuda}: 
(a) {\mbtrack}: each bunch is assigned to one CPU core, (b) {\mbtrackcuda}: all the bunches are on the GPU. 
The simulations are parallelized over macro-particles in one bunch.}
\end{figure}

The main difference between the two versions is on which device the operations of the macro-particles would 
be carried out. In the original {\mbtrack}, apparently, all the operations are carried out by CPU. However, 
in the {\mbtrackcuda} code, the coordinates of the macro-particles are all stored on the GPU. The operations 
on the macro-particles are performed also on the GPU, which means the CPU is used only to control the flow of the 
simulation and to write the output data. Due to the different architectures of CPU cluster and GPU, the parallelization 
models used in the two versions are different, as shown in Figure~\ref{fig:mbtrack-parallelize}. {\mbtrack} parallelizes 
the simulations over bunches with each process handling one bunch, while the macro-particles in one bunch are tracked in series. 
On the other hand, the {\mbtrackcuda} parallelizes the simulations over macro-particles in one bunch. Different bunches 
are tracked in series. 

\begin{figure}[htb]
\centering
\includegraphics[width=\linewidth]{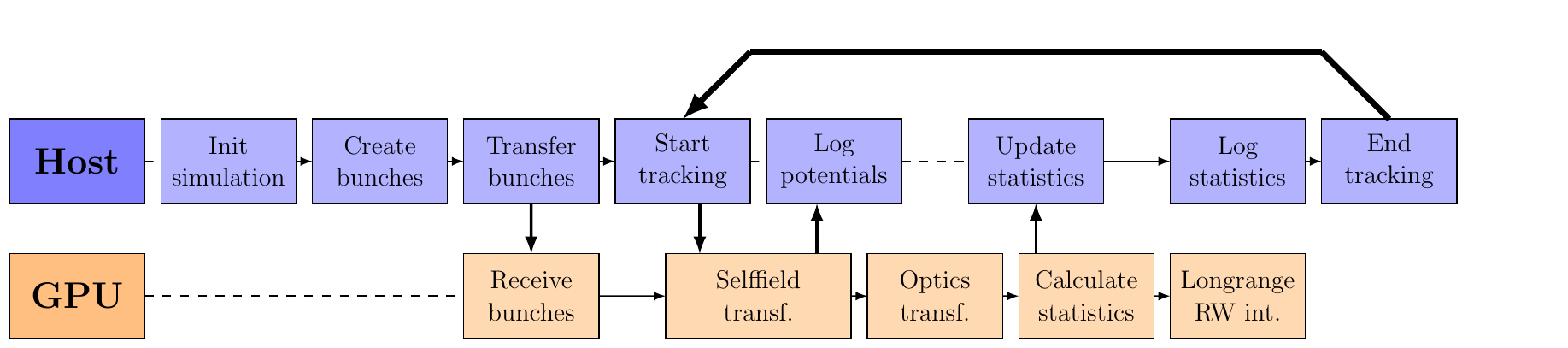}
\caption{\label{fig:mbtrack-flow}Flow diagram of the {\mbtrackcuda} code.}
\end{figure}

To create {\mbtrackcuda},  CUDA kernels have been written to perform all the transformations that are implemented in 
the original {\mbtrack}. In addition, the statistics calculations are also performed on the GPU to avoid transferring 
the macro-particles' coordinates to the CPU side. The flow diagram of {\mbtrackcuda} is shown in Figure~\ref{fig:mbtrack-flow}. 
The flow diagram shows the tasks executed on the host and the kernels launched on the GPU side.
The transformations performed by {\mbtrack} and the implementation of CUDA kernels for {\mbtrackcuda} are described 
in detail in the rest of the section.

\subsection{The basic single-particle transformations}
Every particle in {\mbtrack} is represented by a 6 dimensional vector:

\begin{equation*}
(x, x, y, y', \tau, \delta)
\end{equation*}

\noindent where $x$ and $y$ are the horizontal and vertical positions, while $x'$ and $y'$ are transverse momenta at 
longitudinal position s. Parameter $\delta={\Delta E}/{E_0}$ describes the energy deviation relative to the reference particle, 
and the longitudinal coordinate $\tau$ is the arrival time with respect to the reference particle \cite{Skripka2016221}.

At each turn this transformation computes the energy deviation for each particle:

\begin{equation}
\delta_{i+1} = \delta_{i} + \epsilon_{i} - \frac{U_{rad}}{E_0}
\end{equation}

\noindent where $\epsilon_{i}$ is the relative energy gain in the RF cavities, $U_{rad}$ is the average energy loss per 
turn due to the synchrotron radiation (SR), and $E_0$ is the reference energy. After energy deviation is computed, the longitudinal 
coordinate is updated as follows:

\begin{equation}
\tau_{i+1} = \tau_i + \delta_i T_0 \alpha_c
\end{equation}

\noindent where $\alpha_c$ is the momentum compaction factor.

In transverse planes the particles in the beam perform betatron oscillations. These oscillations are described using Twiss 
parameters $\alpha_{x,y}, \beta_{x,y}, \gamma_{x,y}$, which describe the beam shape, size and orientation, and a phase advance 
per turn $\Psi_{xy}$ \cite{Skripka2016221}. For particles with non-zero energy deviations the phase advance can be calculated by:

\begin{equation}
\Psi_{xy} = \Psi_{x0y0}(1 + \xi_{xy}\delta)
\end{equation}

\noindent where $\xi_{xy}$ is the chromaticity. Given the presence of the horizontal dispersion $D$, the transformation in 
transverse planes are expressed by the transfer matrices:

\begin{equation}
  \begin{aligned}
  \begin{pmatrix} x \\ x' \\ \delta \end{pmatrix} & =
  \begin{pmatrix}
    \cos\Psi_x+\alpha\sin\Psi_x & \beta_x\sin\Psi_x & D \\
    -\gamma_x\sin\Psi_x & \cos\Psi_x -\alpha\sin\Psi_x & D'\\
    0 & 0 & 1
    \end{pmatrix}
  \begin{pmatrix} x \\ x' \\ \delta \end{pmatrix} \\
  \\
  \begin{pmatrix} y \\ y' \end{pmatrix} & =
  \begin{pmatrix}
    \cos\Psi_y+\alpha\sin\Psi_y & \beta_y\sin\Psi_y  \\
    -\gamma_y\sin\Psi_y & \cos\Psi_y -\alpha\sin\Psi_y
    \end{pmatrix}
  \begin{pmatrix} y \\ y' \end{pmatrix} \\
  \end{aligned}
\end{equation}

Changes in the beam energy are also caused by quantum excitation and radiation damping shown in the equations below.

\begin{equation}
  \begin{aligned}
  \tilde{\delta}_{i+1} = & \delta_{i+1}(1-D_E) + \sigma_E \sqrt{2D_e}\delta_{rand}\\
  \tilde{x}_{i+1} = & x_{i+1} + \sigma_x \sqrt{D_x} x_{rand} \\
  \tilde{x'}_{i+1} = & x'_{i+1}\frac{1+\delta_{i+1}}{1+\delta_{i+1}+\epsilon_{i+1}} + \sigma_{x'}\sqrt{D_x}x'_{rand}
  \end{aligned}
\end{equation}

\noindent where coefficients $D$ and $\sigma$ correspond to the synchrotron radiation damping time and bunch energy spread, 
while $\delta_{rand}, x_{rand}$ and $x'_{rand}$ are random numbers from normal distribution with unit standard deviation \cite{Skripka2016221}.

The application launches one CUDA kernel that performs these transformations for every bunch. Since calculations for each 
particle are independent, one thread per particle is created inside the kernel. The random numbers needed for the calculations 
of the radiation damping and quantum excitation are generated using the NVIDIA cuRAND library. Shared memory is used to hold 
the data for additional harmonic cavities, since they are shared by all the threads, and shared memory usage allows to improve 
the load time from global memory.

\subsection{Treatment of the bunch-wake interactions}

The simulations in {\mbtrack} can include resistive-wall effects, arbitrary number of resonators, and purely resistive and inductive 
components, all contributing to the total wake \cite{Skripka2016221}. The macro-particles in each bunch are grouped into 
cells (bins) depending on their longitudinal position. The wake functions are calculated corresponding to ensemble of resonators. 
Each turn, the excitation of this wake on each bin is calculated and the resulting kick is given to every particle in the bin \cite{mbtrackmanual}.

The wake function is calculated for each resonator and summed to form the total wake. Additionally resistive-wall effects 
are also added to the same wake. In the longitudinal and transverse planes the transformation is expressed as change in particle 
energy (horizontal and vertical planes are treated identically):

\begin{equation}
  \begin{gathered}
    \Delta\delta_j = \frac{q_j V(\tau_j)}{E_0} \\
    \Delta x_j' = \frac{q_j}{E_0} \displaystyle\sum_{k=0}^{j-1}q_k D_p(\tau_k) W^{\perp}(\tau_j - \tau_k)
  \end{gathered}
\end{equation}

\noindent where $V(\tau_j)$ is the wake voltage induced by this bunch at bin $\tau_j$, $W^{\perp}$ is the transverse 
wake function, $q_{j,k}$ is the total charge in the bins $j$ and $k$, and $D_p(\tau_k)$ is the dipole-moment of the bunch 
at position $\tau_k$. Detailed information about how $V(\tau_j)$ and $W^{\perp}$ are calculated can be found in \cite{Skripka2016221}.

The transformation is carried out step-by-step. Each step begins by assigning each macro-particle to a bin (mesh cell). 
In the same step, the calculations of different macro-particles are independent. Therefore, this process can be parallelized 
over the number of macro-particles in the bunch. A kernel is launched to put all the macro-particles to their own bins 
using the strategy that each thread handles one macro-particle. The bin number is saved in a temporary memory reserved in 
the beginning of the calculations and is reused between bunches.

After each macro-particle has been assigned to a bin, another kernel is launched to count the number of macro-particles 
in each bin and the dipole-moments $D_p$ for horizontal and vertical planes. This kernel is also launched with one thread 
per macro-particle. However, since there are a number of macro-particles per bin, the atomic operations are needed to sum 
up all the number of macro-particles in the bins.

After the information of the macro-particles per bin is known, a kernel is launched to find the bins holding the minimum 
and maximum number of macro-particles. Since this operation requires communication between threads, it's not suitable for 
parallelization. Therefore, this operation is performed serially on the GPU. One block with multiple threads are launched on the 
GPU, the multiple threads parallelize the loading of the data from the slow GPU memory to the faster shared memory.  And then, 
one thread searches for the first and last bins that contains macro-particles. Since we do not need to loop over all the cells, 
this serialization of the kernel will not cause a bottleneck for the simulation.

Once the wake potentials are constructed, a kernel is called to carry out the bunch-wake interactions. This kernel parallelizes 
the simulation over the macro-particles in the bunch, and launches one thread for each macro-particle. Shared memory is used to 
store the data that is frequently reused by the kernel, which can minimize the loads from global memory.

After the wake potentials have been applied to the macro-particles, the long-range resistive-wall effects are calculated and 
applied to the transverse planes as described in \cite{Skripka2016221}. Since long-range resistive-wall effects require statistics 
information about previous bunches, these calculations are launched only when wake potential effects are calculated for all the 
bunches and the statistics have been updated.

\subsection{Statistics calculations}

The statistics of the bunches are calculated turn-by-turn in order to save the temporary information of the bunches during the 
simulations. To calculate the statistics, such as the mean values and the RMS values of the six dimensional coordinates of each bunch, 
is very time consuming mainly because of the relatively large number of macro-particles and the communication time among different threads. 
Furthermore, since the calculations of the statistics will be carried out on the GPU, the calculated statistics data need to be transferred to the CPU 
in order to log the statistics to file at every turn,
which is also computationally intensive. To calculate the average values, the Thrust libraries \textit{reduce} function is used to calculate the sums of 
position and momentum in each dimension. After \textit{reduce} is performed, data will be sent to the CPU. To compute the standard deviation, 
the Thrusts \textit{transform\_reduce} function is used. 

Since each macro-particle in the bunch in {\mbtrack} is represented as a structure of 6 variables, custom operators are defined to 
perform reduction and transformations correctly on the arrays of particles. The statistics are calculated for each bunch as well as averaged 
for all the bunches in the simulations. Only the calculations for the individual bunches are performed on the GPU while the rest are carried out on the CPU.

\section{Benchmarks of the {\mbtrackcuda}}
\label{sec:CodeBenchmark}

\begin{figure}[htb]
\centering
\subfigure[]{\includegraphics[width=0.45\linewidth]{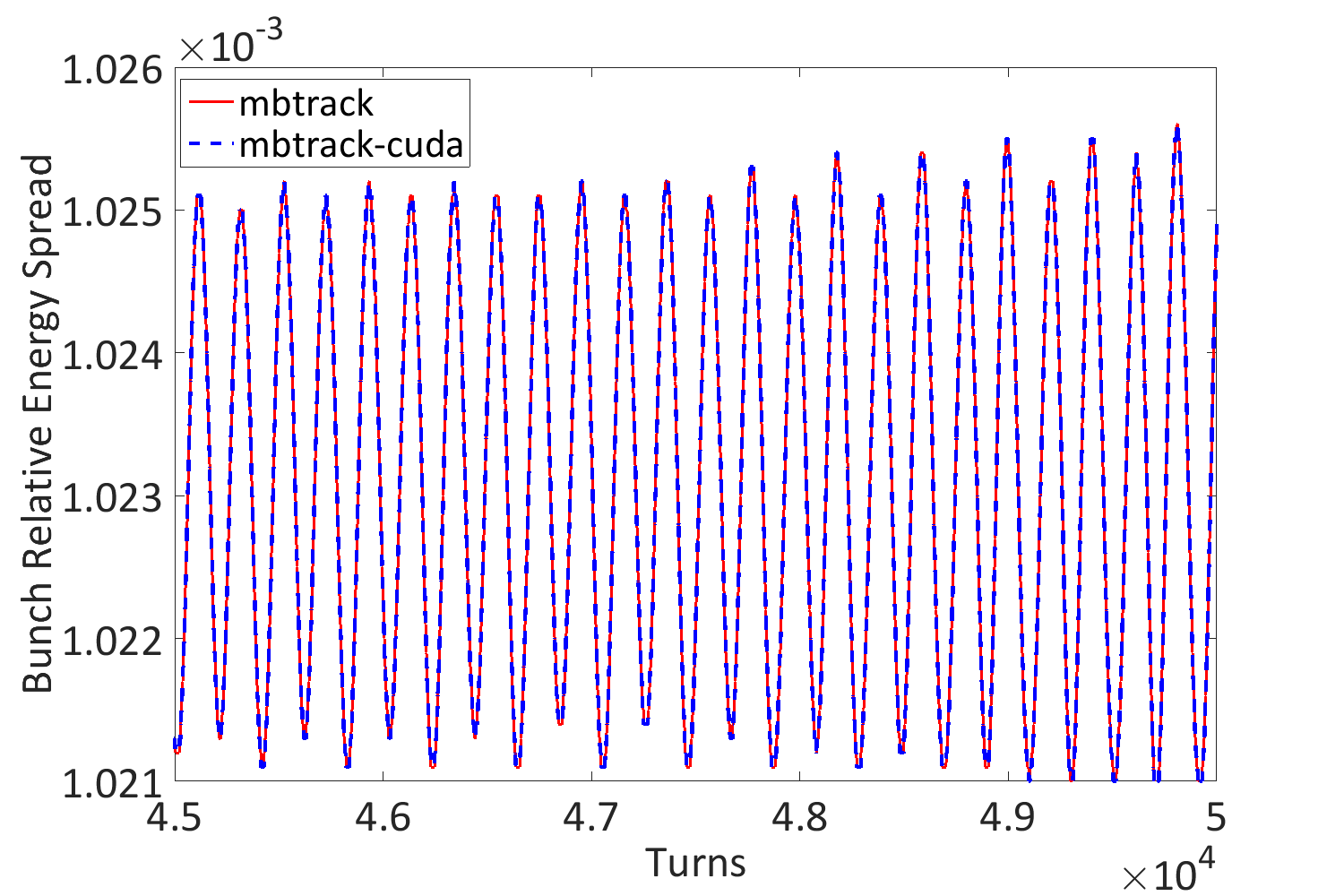}}
\subfigure[]{\includegraphics[width=0.45\linewidth]{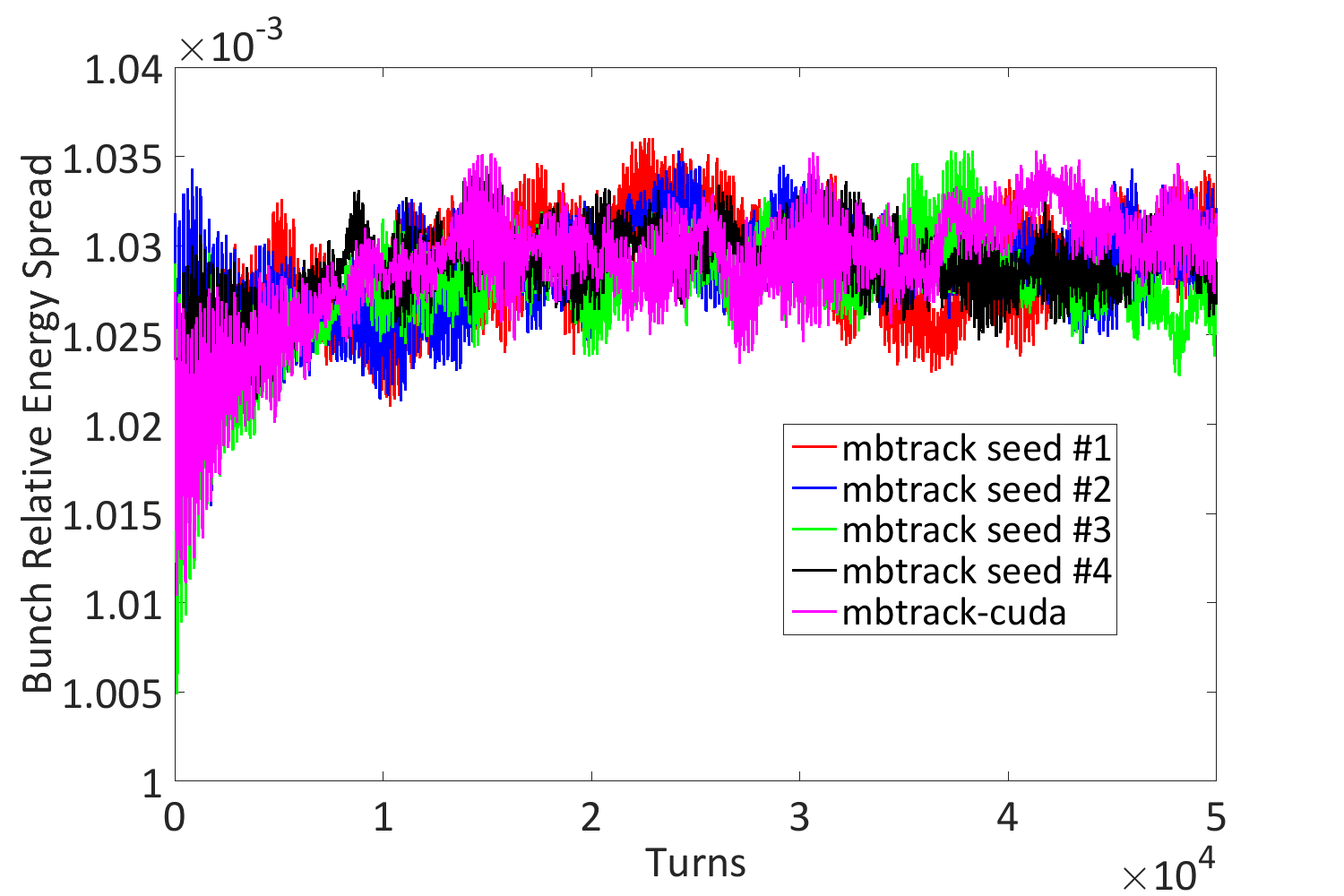}}
\caption{\label{fig:testTracking} (a) Energy spread vs. turns without synchrotron radiation effects; (b) Energy spread vs. turns 
with synchrotron radiation effects. When turning on the synchrotron radiation effects, four different seeds are used to initialize 
the random number generator in {\mbtrack}.}
\end{figure}

Before using {\mbtrackcuda} in our studies, it was essential to carry out systematic benchmarks of the code, in which, the tracking module is the 
first thing to be tested. In the first test, the same group of macro-particles are tracked for 50000 turns (about seven longitudinal damping time), while 
turning off the impedance induced collective effects in both codes. The lattice 'dc12c', which was an option of the SLS-2 storage ring, is used in 
the simulations. The main parameters of the 'dc12c' lattice are listed in Table~\ref{tab:dc12c}. The bunch energy spread is monitored, 
as shown in Figure~\ref{fig:testTracking}. 

\begin{table}[htbp]
   \centering
   \caption{the Main parameters of the 'dc12c' lattice}
   \label{tab:dc12c}
   \begin{tabular}{cc}
     \hline
     Parameters & Values \\
     \hline
     Circumference $C_{\rm ring}$ & 290.4~m\\
     Beam Energy $E_0$ & 2.4~GeV \\
     Radiation Energy Loss per Turn $U_0$ & 569.604~keV \\
     Momentum Compaction Factor $\alpha_c$ & $-1.356\times10^{-4}$\\
     Betatron Tune $\nu_x~\backslash~\nu_y$ & 37.221 $\backslash$ 10.323\\
     Natural Chromaticity $\xi_x~\backslash~\xi_y$ & -66.591 $\backslash$ -40.445 \\
     Transverse Emittance $\epsilon_x~\backslash~\epsilon_y$ & 138.55~pm $\backslash$ 0.12~pm \\
     Harmonic Number h & 484 \\
     Peak RF Voltage $V_{\rm RF}$ & 1.4~MV \\
     Synchrotron Tune $\nu_s$ & 2.48$\times10^{-3}$ \\
     \hline
\end{tabular}
\end{table}

To eliminate the influence of the different random number generators in the two different platforms, we first turned off the 
synchrotron radiation effects in the tracking. The bunch energy spread in the last 5000 turns are shown in the 
Figure~\ref{fig:testTracking} (a) as an example. We can find from this figure that the results produced from both codes 
agree perfectly with each other, which demonstrates the reliability of the tracking module in {\mbtrackcuda}.

However, when the synchrotron radiation effects are turned on, a small discrepancy between the two codes can be observed, as shown 
in the Figure~\ref{fig:testTracking} (b). We believe that this small discrepancy is because of the random numbers when considering 
synchrotron radiation. We then vary the seeds of the random number generator in {\mbtrack}. the results corresponding to four 
different seeds are also shown in Figure~\ref{fig:testTracking} (b). As we can see, the discrepancy between the two codes is 
comparable with the cases when changing the seeds of the random number generator in {\mbtrack}, meaning that the discrepancy 
between the two codes is dominated by the random number generators. We believe that the above mentioned results demonstrate the good 
agreement between the two codes. 

\begin{figure}[htb]
\centering
\subfigure[]{\includegraphics[width=0.45\linewidth]{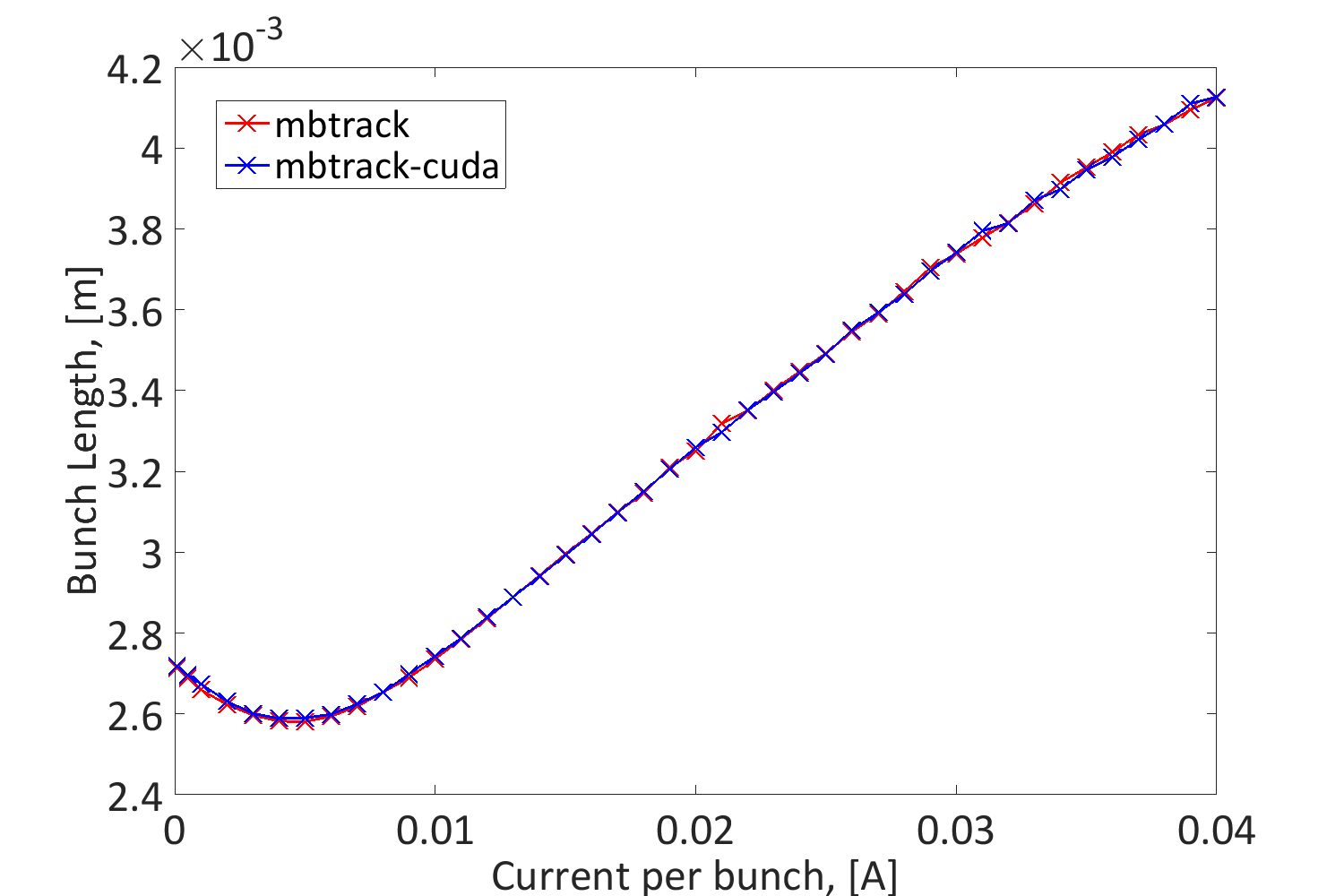}}
\subfigure[]{\includegraphics[width=0.45\linewidth]{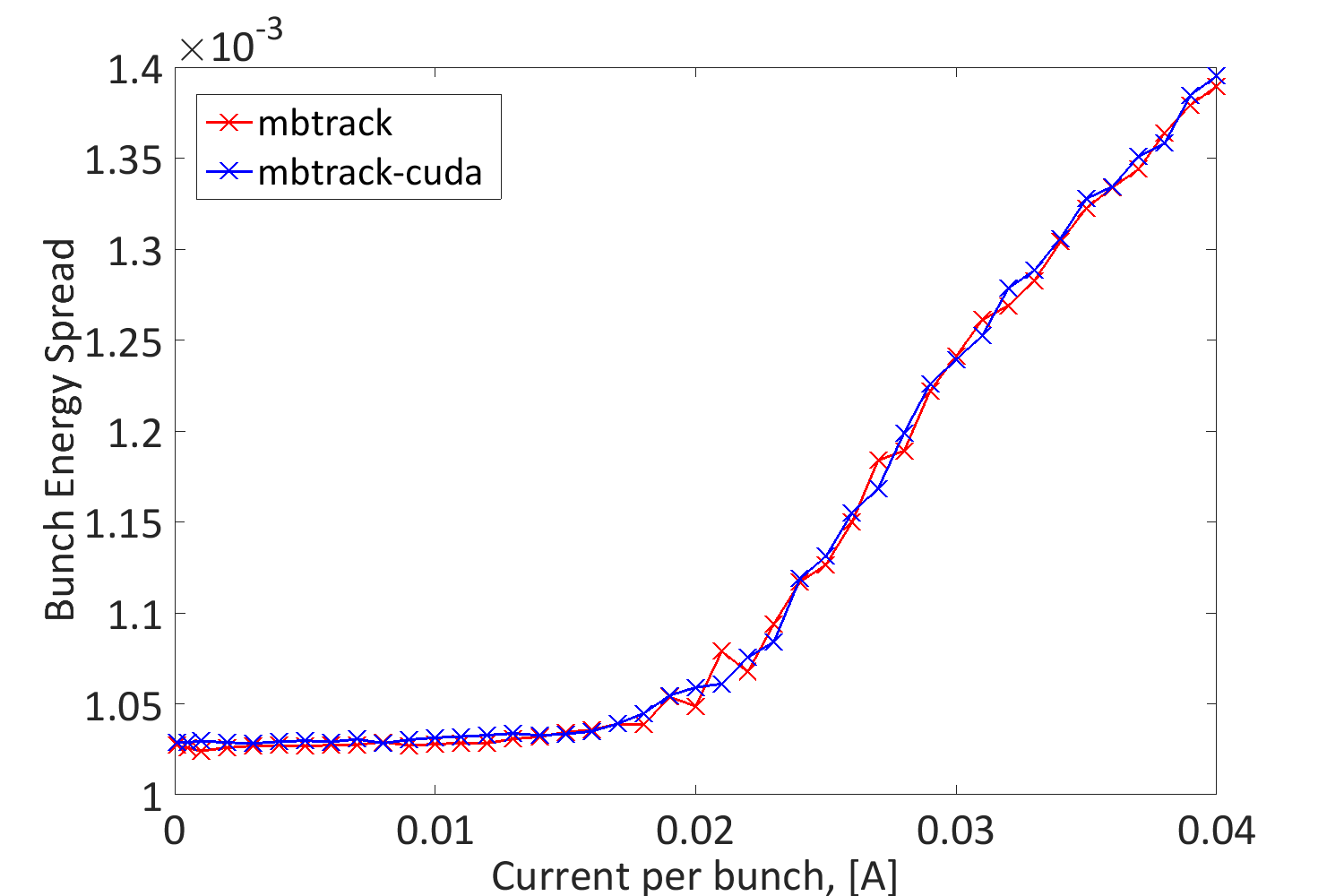}}
\caption{\label{fig:testWake} Comparison of {\mbtrack} and {\mbtrackcuda}. (a) 'Equilibrium' bunch lengths vs. single-bunch currents; (b) 'Equilibrium' energy spread vs. single-bunch currents.}
\end{figure}

After testing the tracking module of {\mbtrackcuda}, we benchmark the beam-impedance interactions. 
The longitudinal short-range resistive-wall (RW) wake field is used in the test. Here, we assume that the copper vacuum chamber with 10~mm 
inner radius is along the whole ring. Neither geometric wake nor RW wake of any other component is included. 
The macro-particles are tracked for 50000 turns as well in this test. The 'equilibrium' 
bunch length and energy spread can be calculated by averaging the tracking data in the last 5000 turns. By varying the single-bunch current, the plots in 
Figure~\ref{fig:testWake} are obtained. 

Figures~\ref{fig:testWake} (a) and (b) show how the 'equilibrium' bunch length and energy spread vary when increasing the single-bunch 
current, respectively. The 'equilibrium' bunch length, shown in Figure~\ref{fig:testWake} (a), first reduces when the 
single-bunch current goes up from zero, and then, increases with the rising bunch current. The bunch shortening effect at low current is 
due to the negative momentum compaction factor of the used lattice and an inductively dominated wake. The 'equilibrium' energy spread keeps almost constant below about 15~mA 
and keeps growing above this current, which is the microwave instability threshold. Furthermore, both Figure~\ref{fig:testWake} (a) and (b) 
indicate clearly the desired good agreement between the two codes.

After the validity test of the {\mbtrackcuda} code, we carried out performance tests. The preliminary performance tests were performed in a 
system equipped with 2x Intel E5-2609 v2 CPUs (2x4 CPU cores, 2x4 threads maximum) and an NVIDIA Tesla K40C graphics card (2880 CUDA cores). 
The two example lattices integrated in {\it mbtrack} source code, which are SOLEIL lattice and MAX~IV 3~GeV ring lattice, are used in our preliminary 
performance tests, respectively. In the tests, each bunch, consisting of 100,000 macro-particles, were tracked for 10,000 turns. 

In the test using SOLEIL lattice, the basic optics transformations and long range RW effects are implemented. The tracking was performed for longitudinal 
and horizontal planes. Since the random number generator will be called every turn if the synchrotron radiation is included in the simulation, 
we manually turn the synchrotron radiation effects on and off to find the influence of generating random numbers on the simulation time in both codes. 
The results in Table~\ref{tab:mbtrack-soleil} show the full execution time of the simulations including the input and output operations. 

\begin{table}[ht]
\centering
\caption{\label{tab:mbtrack-soleil}Comparison of the computing time using SOLEIL lattice.}
\footnotesize{
\begin{tabular}{|c|c|c|c|c|}
\hline
\multirow{2}{*}{Bunches} & \multicolumn{2}{c|}{No SR effects} & \multicolumn{2}{c|}{With SR effects}\tabularnewline
\cline{2-5}
& {\mbtrack} & {\mbtrackcuda} & {\mbtrack} & {\mbtrackcuda}\tabularnewline
\hline
1 & 116~s & 44~s & 942~s & 44~s\tabularnewline
\hline
2 & 117~s & 65~s & 940~s & 65~s\tabularnewline
\hline
3 & 122~s & 86~s & 970~s & 86~s\tabularnewline
\hline
10 & 457~s & 231~s & 2059~s & 360~s\tabularnewline
\hline
138 & - & \textcolor{black}{2992~s} & - & \textcolor{black}{2982~s}\tabularnewline
\hline
416 & - & \textcolor{black}{9452~s} & - & \textcolor{black}{9461~s}\tabularnewline
\hline
\end{tabular}
}
\end{table}

In the test using the MAX~IV 3~GeV ring lattice, one passive third-harmonic cavity and one broad-band resonator are included. 
The transformations are performed only in the longitudinal plane. The results are shown in Table~\ref{tab:mbtrack-maxiv}.

\begin{table}[ht]
\centering
\caption{\label{tab:mbtrack-maxiv}Comparison of the computing time using MAX~IV 3~GeV ring lattice.}
\footnotesize{
  \begin{tabular}{|c|c|c|c|c|}
    \hline
    \multirow{2}{*}{Bunches} & \multicolumn{2}{c|}{No SR effects} & \multicolumn{2}{c|}{With SR effects}\tabularnewline
    \cline{2-5}
    & {\mbtrack} & {\mbtrackcuda} & {\mbtrack} & {\mbtrackcuda}\tabularnewline
    \hline
    1 & 82~s & 33~s & 357~s & 33~s\tabularnewline
    \hline
    2 & 81~s & 59~s & 355~s & 60~s \tabularnewline
    \hline
    3 & 87~s & 86~s & 362~s & 86~s\tabularnewline
    \hline
    10 & 273~s & 249~s & 856~s & 356~s\tabularnewline
    \hline
    58 & - & \textcolor{black}{1668~s} & - & \textcolor{black}{1689~s}\tabularnewline
    \hline
    176 & - & \textcolor{black}{7514~s} & - & \textcolor{black}{7511~s}\tabularnewline
    \hline
  \end{tabular}
}
\end{table}

One can find from the validation tests that the implementation of synchrotron radiation results in remarkable increase 
of the computation time when using {\mbtrack}, which is due to the call of the random number generator every turn. However, the 
SR effect has little influence on the computation time by {\mbtrackcuda}. This fact is because the well optimized random number 
generator on GPU platform was used in {\mbtrackcuda}. We benefit from the highly parallelized architecture of GPU here in the 
generation of random numbers. 

Furthermore, we can find from Table~\ref{tab:mbtrack-soleil} and Table~\ref{tab:mbtrack-maxiv} that for this computer, no results can be generated by 
{\mbtrack} when the number of bunches is significantly more than the number of CPU cores (8 CPU cores). Meanwhile, the {\mbtrackcuda}
still manages to run the multi-bunch simulations, even the full-ring multi-bunch simulations, in reasonable time. This fact shows the 
significance of our development.  

We also carried out the tests of computation time using the above mentioned SLS-2 lattice 'dc12c'. In these simulations, 
each bunch consisted of 100,000 macro-particles and was tracked for 20,000 turns. Both synchrotron radiation damping and 
quantum excitation effects were enabled. Here, we carried out the test in three different systems. The first 
system was the same stand-alone workstation mentioned above, which has 2x Intel E5-2609 v2 processors (2x4 CPU cores). However, the above mentioned NVIDIA 
graphics card was moved to the second system, which was a multi-core high-performance workstation equipped with 2x Intel E5-2697 v4 
processors (2x18 CPU cores, hyper-threading enabled). The third system was a cluster equipped with 32 Intel XEON GOLD 6140 processors (18 CPU cores, 
hyper-threading disabled). We managed to fully parallelize the simulations in the third system.  
The computing time of the different tests is shown in Table~\ref{tab:mbtrack-benchmark}.

\begin{table}[htbp]
  \centering
  \caption{Benchmarks of {\mbtrack} and {\mbtrackcuda} on CPU (8-core), CPU (36-core), cluster (576-core CPU), and GPU (2880 CUDA cores).}
  \label{tab:mbtrack-benchmark}
  \footnotesize{
    \begin{tabular}{|c|c|c|c|c|}
      \hline
      \multirow{2}{*}{Number of Bunches} & \multicolumn{4}{c|}{Computing Time}\tabularnewline
      \cline{2-5}
      & 8-core CPU & 36-core CPU & cluster & GPU\tabularnewline
      \hline
      1 & 810~s & 549~s & 232~s & 93~s\tabularnewline
      \hline
      2 & 814~s & 558~s & 236~s & 157~s\tabularnewline
      \hline
      3 & 832~s & 562~s & 240~s &  221~s\tabularnewline
      \hline
      10 & 3596~s & 577~s & 266~s &  546~s\tabularnewline
      \hline
      20 & - & 757~s & 312~s &  \textcolor{black}{1006~s}\tabularnewline
      \hline
      121 & - & 2883~s & 922~s & \textcolor{black}{5860~s}\tabularnewline
      \hline
      363 & - & 10931~s & 3165~s & \textcolor{black}{18754~s}\tabularnewline
      \hline
    \end{tabular}
  }
\end{table}

The Table~\ref{tab:mbtrack-benchmark} shows significant difference of the computation time when running {\it mbtrack} in the above mentioned CPUs. This phenomenon is mainly because of the different performance of the CPU cores and the different number of CPU cores. It's interesting
to point out that the full-ring multi-bunch simulation can be carried out in a multi-core (36 cores with hyper-threading enabled) stand-alone workstation. 
However, the computation time will be remarkably longer than in the cluster mainly because of the heavy overloading of the CPU cores and the lower performance 
of each core. 

The present version of {\mbtrackcuda} shows higher speed when the number of bunches is fewer, e.g., single-bunch simulations. However, when running the 
simulations with more and more bunches, the {\mbtrackcuda} becomes slower. For instance, it's about 6 times slower to carry out 363 bunches 
simulation by the {\mbtrackcuda} code in the above mentioned workstation than carrying out the same simulation in the above mentioned cluster by {\mbtrack}.
The significant degradation of the performance as the number of bunches increases, is mainly because the present {\mbtrackcuda} code has to calculate the 
different bunches in series. However, it shows the potential to accelerate the simulations in the future.

\section{Simulations of the longitudinal coupled-bunch instability for SLS-2 by {\it mbtrack-cuda}}
\label{sec:CodeSimulations}

In this section, we present our implementation of the developed {\mbtrackcuda} code in the study of the longitudinal 
coupled-bunch instability in SLS-2 (using 'dc12c' lattice). The 500~MHz ELETTRA-type RF cavities, used in SLS \cite{SLSCAVITYPAC99}, are considered to be reused in SLS-2. 
We therefore use the same Higher Order Modes (HOMs) parameters of the SLS cavity in the following simulations.

\begin{figure}[ht]
\centering
\includegraphics[width=0.9\textwidth]{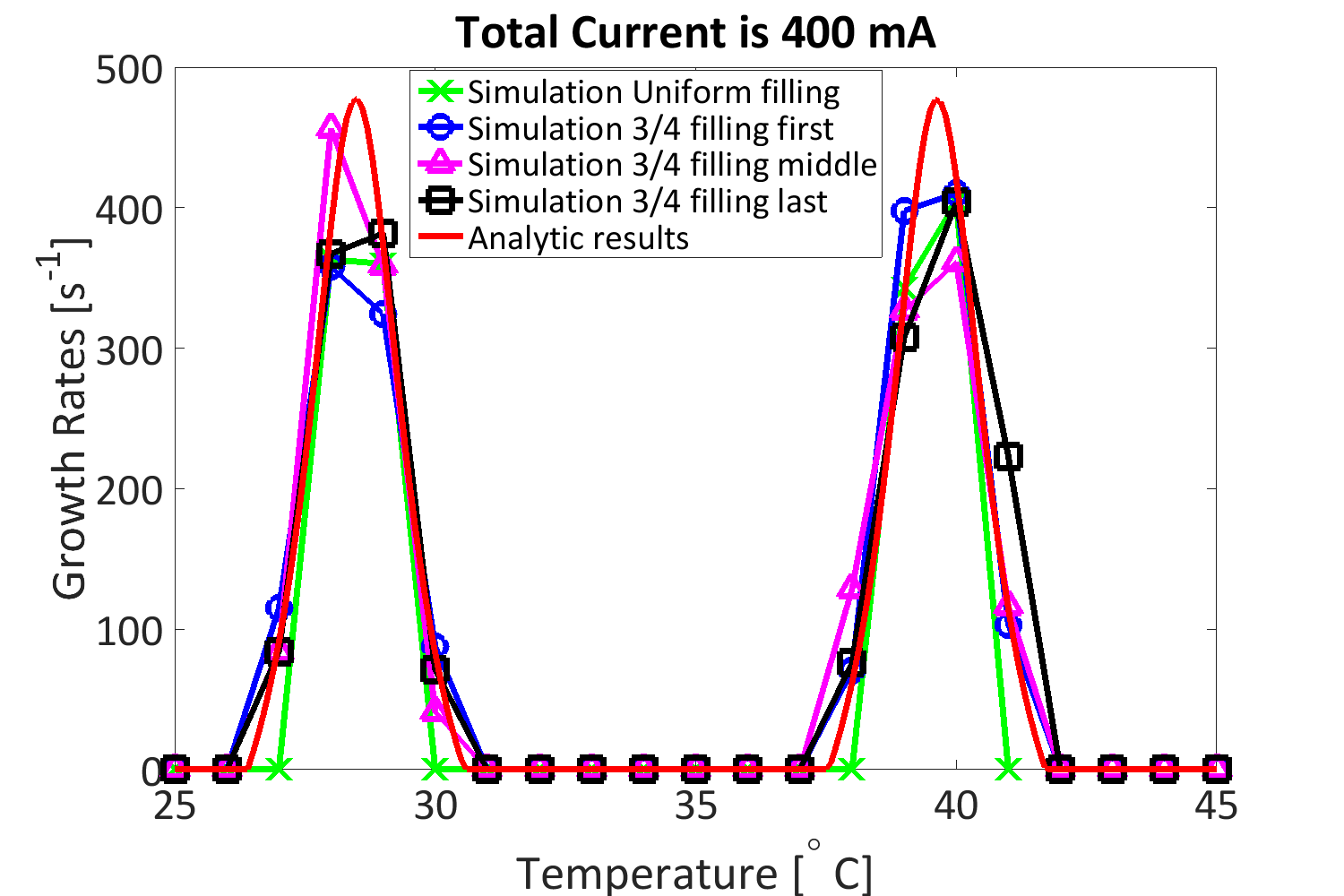}
\caption{\label{fig:LCBICompare} The comparison of the growth rates of the longitudinal coupled-bunch instability driven by the longitudinal 
Higher-Order Mode L5 of the Cavity \#3. The red curve is the analytic estimation of the growth rate induced by the mode L5; 
the green curve with 'cross' markers show the growth rates extracted from the simulation data using the uniform filling pattern; 
the blue curve with 'circle' markers, the magenta curve with 'triangular' markers, and the black curve with 'square' markers correspond to 
the growth rates of the first bunch, the middle bunch, and the last bunch in the bunch train in the \sfrac{3}{4} filling pattern, respectively. 
In the \sfrac{3}{4} filling pattern, 363 continuous buckets out of 484 buckets are filled identically. To make a fair comparison, the total current of 
400~mA is used in both the analytic estimation and the simulations.}
\end{figure}

ELETTRA type cavities utilize the temperature 
of cooling water and the plunger tuner as the two main parameters to tune the resonant frequencies of all the modes. 
To avoid the longitudinal coupled-bunch instability, 
the temperature of the cooling water should be adjusted to the values in the 'stable windows'~\cite{Svandrlik:1995zi,Svandrlik:1997zz}. 
Using the L5 mode (resonant frequency 
1606.862~MHz at $45^\circ$C \cite{TALKPAOLO}) in the cavity \#3 as an example, the analytical estimation of the growth rates for M equal bunches, for longitudinal 
coupled-bunch mode $\mu$, shown by Eq.~(\ref{eq:LCBI}) \cite{NGBOOK}, have been carried out under the assumption of uniform filling pattern. The result is 
shown by the red curve in Figure~\ref{fig:LCBICompare} for the case of SLS-2 without harmonic cavity. 

\begin{eqnarray}
  \frac{1}{\tau_{1\mu}}&=&\frac{\eta e^2 N_b M \omega_r}{2\beta^2E_0T_0^2\omega_s}\cdot\nonumber\\
  &&\left[\mathscr{R}{\it e}~ Z_0^\parallel\left(qM\omega_0+\mu\omega_0+\omega_s\right) - \mathscr{R}{\it e}~ Z_0^\parallel\left(q'M\omega_0-\mu\omega_0-\omega_s\right)\right]
  \label{eq:LCBI}
\end{eqnarray}

The above mentioned analytical method uses an assumption of uniform filling of all the buckets, which is usually not the case in 
the real operation of the storage rings in synchrotron light sources. As mentioned above, we propose that 390 identical bunches are 
filled continuously in the SLS-2 storage ring. The goal of the simulation is to see, whether the stable temperature windows shift 
because of the nonuniform filling. We therefore simulate the influence of the L5 mode using the uniform filling pattern and the \sfrac{3}{4} filling 
pattern, respectively. To compare with the analytical estimations, we plot the growth rates of the bunches under different conditions 
in Figure~\ref{fig:LCBICompare}. The green curve with the 'cross' markers shows the simulation results at the uniform filling pattern. 
Meanwhile, the information of the first bunch, the middle bunch, and the last bunch in the bunch train of the continuous \sfrac{3}{4} filling pattern 
are all shown in the same figure. 

The analytic method is able to provide a satisfactory good prediction of the 'stable window'. By comparing the 
simulation results under the assumption of uniform filling pattern and the \sfrac{3}{4} filling, we could find that the 
resulting 'stable window' is almost the same. This study provides more confidence to the analytic estimations of the longitudinal coupled-bunch instability.

\section{Conclusions}
\label{sec:Conclusion}

We present the development of the {\mbtrackcuda} code, using the GPU computing technology, in this paper. The heaviest computations 
are carried out by GPU in this code. {\mbtrackcuda} allows to run multi-bunch simulations in one stand-alone workstation with a scientific 
graphics card in an acceptable time. Therefore, this code reduces the requirement of the large scale clusters, which is usually expensive 
in construction and operation. 

In the present version of {\mbtrackcuda}, the performance still needs improvements since it's still slower than running {\mbtrack} 
in not only a big cluster, but also a multi-core high performance stand-alone workstation. However, it shows advantage in the single-bunch 
simulations, which clearly demonstrate the great potential of improvement of {\mbtrackcuda}. 

The {\mbtrackcuda} code has been implemented in the study of the longitudinal coupled-bunch instabilities of the SLS-2 storage ring. 
The longitudinal HOM L5 of the ELETTRA-type 500 MHz cavity is studied both analytically and by simulation. 
By changing the cooling water temperature of the cavity, 
we simulated both the uniform filling pattern and the \sfrac{3}{4} filling pattern at the total current 400~mA, respectively. The simulation results 
agree well with the analytical estimation.

The limiting factor of the {\mbtrackcuda} version is the lack of multi GPU implementation. Since {\mbtrackcuda} shows significant performance 
improvement for a single bunch simulations splitting the bunches among multiple GPUs would allow multi node clusters to take full advantage of 
GPU resources. The multi GPU implementation would also decrease the memory used by a single card thus allowing to run bigger simulations with 
more macro-particles per bunch.

With the advances in GPU technologies the number of CUDA cores keeps increasing and so does the memory speed which should lead to even better 
performance. Since the number of macro particles per bunch is usually very high, {\mbtrackcuda} will be able to take advantage if increasing 
core counts. Adapting to the new VOLTA architecture would certainly decrease the time to solution.

\section{Acknowledgements}

The authors would like to thank Dr. Ryutaro Nagaoka and Dr. Francis Cullinan for their kind support and discussions of {\mbtrack} and  
Dr. Paolo Craievich for providing the information of the HOMs of SLS cavities. The authors would also like to thank Carl Beard for the 
kind English corrections. 

The research leading to these results has received funding from the European Community's Seventh Framework 
Programme (FP7/2007-2013) under grant agreement n.o290605 (PSI-FELLOW / COFUND). The author Haisheng Xu would like to thank the PSI-FELLOW program 
for the support. 

The 576-core cluster used in the code benchmark is operated by the IHEP computing center. The authors would like to thank the staffs in the IHEP 
computing center for their kind support.

\bibliography{mbtrack}{}
\bibliographystyle{unsrt}

\end{document}